# Towards Resolving Software Quality-in-Use Measurement Challenges

[1] Issa Atoum, [2] Chih How Bong, [3] Narayanan Kulathuramaiyer

[1,2,3] Faculty of Computer Science and Information Technology, University Malaysia Sarawak, 94300 Kota Samarahan,Sarawak,Malaysia

[1] atoum@siswa.unimas.my, [2] chbong@fit.unimas.my, [3] nara@fit.unimas.my

## ABSTRACT

Software quality-in-use comprehends the quality from user's perspectives. It has gained its importance in e-learning applications, mobile service based applications and project management tools. User's decisions on software acquisitions are often ad hoc or based on preference due to difficulty in quantitatively measure software quality-in-use. However, why quality-in-use measurement is difficult? Although there are many software quality models to our knowledge, no works surveys the challenges related to software quality-in-use measurement. This paper has two main contributions; 1) presents major issues and challenges in measuring software quality-in-use in the context of the ISO SQuaRE series and related software quality models, 2) Presents a novel framework that can be used to predict software quality-in-use, and 3) presents preliminary results of quality-in-use topic prediction. Concisely, the issues are related to the complexity of the current standard models and the limitations and incompleteness of the customized software quality models. The proposed framework employs sentiment analysis techniques to predict software quality-in-use.

## 1. INTRODUCTION

With thousands of software published online, it is essential for users to find the software that matches their stated or implied needs. Users often seek better software quality. Garvin [1] identified five views/approaches of quality. The nearest definition in this paper is the user based approach definition "meeting customer needs". If the customer is satisfied, then product or service has good quality. It has been implemented in mobile-based applications[2]–[4] and Web applications[5]–[7].

Software quality can be conceptualized from three dimensions; the quality characteristics, the quality model, and software quality requirements. A Quality characteristic is "category of software quality attributes that bears on software quality" [8, p. 9]. Quality requirements are what the user needs in the software such as performance, user interface or security requirements. The quality model is how quality characteristics are related to each other and to the final product quality. Measuring the software quality will check if user requirements are met and decide the degree of quality.

The ISO/IEC 25010:2010 standard (ISO 25010 hereafter), a part of a series known as the Software Quality Requirements and Evaluation (SQuaRE), defines systems' quality as "the degree to which the system satisfies the stated and implied needs of its various stakeholders, and thus provides value" [9, p. 8]. The ISO 25010 has two major dimensions: Quality-in-use (QinU) and Product Quality. The former specifies characteristics related to the human interaction with the system and the latter specifies characteristics intrinsic to the product. QinU is defined as "capability of a software product to influence users' effectiveness, productivity, safety and satisfaction to satisfy their actual needs when using the software product to achieve their goals in a specified context of use" [8, p. 17].The QinU model consists of five characteristics: effectiveness, efficiency, satisfaction, freedom from risk and context coverage. Table 1 illustrates the definition of these characteristics.

**Table 1:** definitions of quality-in-use characteristics as defined by the ISO 25010 standard

| Characteristic | Definition |
|---|---|
| Effectiveness | Accuracy and completeness with which users achieve specified goals (ISO 9241-11). |
| Efficiency | Resources expended in relation to the accuracy and completeness with which users achieve goals (ISO 9241-11). |
| Freedom From Risk | Degree to which a product or system mitigates the potential risk to economic status, human life, health, or the environment. |
| Satisfaction | Degree to which user needs are satisfied when a product or system is used in a specified context of use. |
| Context Coverage | Degree to which a product or system can be used with effectiveness, efficiency, freedom from risk and satisfaction in both specified contexts of use and in contexts beyond those initially explicitly identified. |

### 1.1 Problem Statement

This paper investigates these problems 1) there are many challenges that need to be tackled in order to measure QinU systematically. However, current literature reviews on software QinU does not identify or explain them. To the best of our knowledge this is the first work that specifically identifies and explains the problems towards measuring QinU. 2) Insufficient research on other possible research directions to tackle the first problem. To our knowledge, little work target to resolve QinU problem [10].

### 1.2 Research Contributions
- This paper identifies and explains several problems while measuring software QinU using the standard and customized quality models. This paper is the first that surveys several quality models and explains various challenges to measure QinU. In brief, most of the challenges in ISO standard models are related to the complication and incompleteness of the documents. On the other hand, customized quality models are subject to incomplete models that are designed for their own specific needs.
- Proposes a novel framework to predict software QinU from software reviews. Given the issues related to measuring QinU a framework is presented to resolve these issues. The framework is based on sentiment analysis, an emerging branch of Natural Language Processing. Sentiment analysis or opinion mining targets to analyze textual user judgments about products or services[11], [12]

First major software quality-in-use related models are illustrated. Then, the quality-in-use measurement challenges are explained. Next, a proposed approach is presented and finally, the paper is concluded.

## 2. SOFTWARE QUALITY-IN-USE MODELS

There have been many works in software quality models but to our knowledge, no research has been conducted to summarize the main problems in measuring quality-in-use. Measuring software quality-in-use can be divided in two main frameworks; the standard and customized model frameworks.

### 2.1 Standard Frameworks

There have been many standards that can support software quality, but many of them are rather check list guide. For example, the ISO 9000 family has been criticized in literature not to be used for software [13]. The ANSI/IEEE 730-2002[14] support quality assurance plans. ISO/IEC 15504[15] or as it is known Software Process Improvement and Capability Determination (SPICE), is a set of technical standards documents for the computer software development process and related business management functions. These standards are not designed to address quality-in-use nor specific characteristics of software product quality.

Recently, the Software Product Quality Requirements and Evaluation (SQuaRE) ISO standard series are a result of blending the ISO/IEC 9126 and ISO/IEC 14598 series of standards. The purpose of the SQuaRE series of standards is to assist developing and acquiring software products with the specification of quality requirements and evaluation. From the viewpoint of the stakeholders the quality requirements are specified, the quality of the product is evaluated based on this specification utilizing chosen quality model, quality measurement and quality management process.

| 2503n Quality Requirement Division | 2501n Quality Model Division | 2504n Quality Evaluation Division |
|---|---|---|
| | 2500n Quality Management Division | |
| | 2502n Quality Measurement Division | |
| ISO/IEC 25050 – 25099 SQuaRE Extension Division ||| 

**Fig 1:** Organization of Square series of International Standards

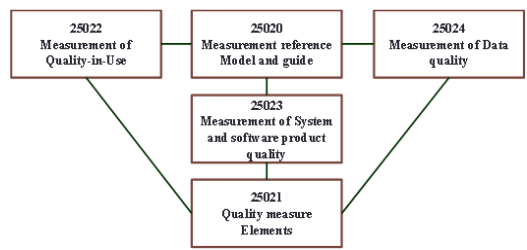

**Fig 2:** Structure of the Quality Measurement division

To measure QinU effectively, five divisions of the SQuaRE series have to be considered the ISO 2502n to ISO 25024 and in line with the ISO 25010 model as shown in Fig. 1. Technically and, more precisely, the QinU Measurement Standard ISO 25022 has to be considered in the context of four other standards: the Measurement Reference Model and Guide ISO 25020; the Measurement of Data Quality 25024, the Measurement of System and Software Product Quality ISO 25023, and Quality Measure Elements Standards ISO 25021. Fig. 2 depicts the relationship between the ISO/IEC 25022 and other ISO/IEC 2502n division of standards.

While these standards provide the freedom of customization, they need careful quality assurance to provide apparent integration between related standards. They also suffer to detail how the customization need to be carried out.

### 2.2 Customized Software Quality Models

Below are some of related models grouped in logical groups.

#### 2.2.1 Hierarchical Models

These models link various quality characteristics together at different levels, which in turn are finally linked with the root product quality. Activity based models can adopt these models to measure software quality. Activity based models provide an interrelation between system properties and the activities carried out by the various stakeholders. Activity based models usually tracks development or testing activities rather than user activities. Famous hierarchical models are McCall's Quality Model [16], Boehm's Quality Model [17], Dromey's Quality Model [18], and The Software Quality In Development (SQUID) approach [19].

### 2.2.2 One-Quality-Aspect Models

These models measure one aspect of software quality. This category includes predictive quality models[20]–[23], quality metrics models[24]–[27] and software reliability models.

Predictive quality models[20]–[23] use the product attributes or its properties and product users to predict the quality of the software. Exploiting ideas in COCOMO[20]the Constructive Quality Model [21], [22] (COQUAMO) helps project managers to manage, assess and predict product during the development lifecycle. Software Metrics Models are concerned with quality evaluation of a specific quality metric, quality assurance, or prediction. Many researchers have shown that these metrics are not certain indicators of faults[28], line code dependent [28] or programming dependent[29][30].Software Reliability Models target to measure the reliability of software systems based on failures intensity and software history profiles. Example of such models are [31], [32]. The application of these models to get quality-in-use is not feasible because the software will be in operation phase for users. I.e., there is neither history nor source code for investigation

### 2.2.3 Provider-Specific Models

There are specific quality models to certain programming language or implementation platforms. FURPS Quality Model was presented by Grady[33] and later extended and owned by IBM Rational Software [34]–[36]. Quamoco Product Quality Model[37] was initially designed for German Software and is a multipurpose quality model based on ISO 25010.

From previously studied models several challenges must be tackled. Section 3 below discusses a list of these challenges.

## 3 QUALITY-IN-USE MEASUREMENT CHALLENGES

Below are major challenges while measuring software quality-in-use in general, measuring quality-in-use using standard frameworks, and measuring quality-in-use using customized models.

### 3.1 General Challenges
#### 3.1.1 Task Measurement

To measure QinU there is a need to agree with the software user on a set of tasks that he/she need to do in order to accomplish a pragmatic goal ("do goals" to achieve the task such as pay a bill). This means the user should be involved in the quality requirements specification which might not be applicable in all times. Other issues related to task measurement embraces the variety of tasks from one software function to another and from software to another. For example, a task to open a file for writing is different than a task of removing special characters from a text file. Worse on this, defining what are the tasks is by itself a major challenge. Hedonic tasks (the "be goals") that imply user satisfaction cannot be specified, thus cannot be measured directly.

#### 3.1.2 The Web Software Development Life Cycle

Users of publicly available online software are never asked to be part of the system development life cycle, but usually the developers and software designer are making assumptions on user needs. In cases where software is designed to be used by global users such as operating systems or antivirus software, then software publishers have to find other ways to collect user needs. However, it might be a disaster when users start using the software. Not because of software bugs, but because users are not satisfied. Users need to see software doing what they were thinking of without draining their mind with all the lifecycle of the software.

#### 3.1.3 Dynamic Customer Needs

Customer needs are dynamic and they can change from time to time, so quantitative measures might not be suitable. Ishikawa [38] states that "International standards established by the International Organization for Standardization (ISO) or the International Electro technical Commission (IEC) are not perfect. They contain many shortcomings. Consumers may not be satisfied with a product which meets these standards. Consumer requirements change from year to year and even frequently updated standards cannot keep the pace with consumer requirements".  These needs are usually resolved by building new software versions, however software might get complicated or buggy due to extra feature added that were not planned ahead. If users are involved ahead of time, these needs will be incorporated. Therefore, this problem returns us to the first and second unsolved issues above.

### 3.2 Challenges Related to Standard Quality Frameworks
#### 3.2.1 Quality Models Critiques

There are problems that are intrinsic to quality models. In a comprehensive study of software quality models,[39] identified critiques to many software quality models; they are unclear of their purposes, not satisfying users of how to use the quality models, and there is no uniform terminology between different models. Masip et

al.[40]stated that user experience is implied in ISO 25010 but is not defined.

### 3.2.2 Evaluation Requirements

Looking into the mathematical formulas of quality-in-use in the ISO 25022 standard and the proposed methods to measure quality-in-use, quality managers find it a hard job. For example, to measure the effectiveness; task completion, task effectiveness and error frequency has to be calculated. Moreover, Integrating related quality processes of various standards (Fig. 1, Fig. 2) is a problem for quality engineers. The reason behind this problem is the need of experienced engineers given limited information in the standard models on how to customize them, especially for small sized companies. In an extension to the ISO 25010 Lew et al. [41] suggest adding data quality inside the ISO 25010 instead of being separate. Monitoring user actions or usage statistics to measure quality-in-use are not enough. A wide range of measuring methods needs an acceptable level of experts in each domain as shown in ISO 25022 standard.

### 3.2.3 QinU Environmental Factors

While quality-in-use model tries to measure the human computer system interaction there are many factors that affect quality-in-use according to the ISO QinU model: the information system, target computer system, target software, target data, usage environment, and user type (primary, secondary, or indirect user). Measuring or estimating these factors is a complex process.

### 3.3 ChallengesRelated to Customized Models
### 3.3.1 Limitation of Quality-in-use Models

Although there are many software quality models such as McCall, Boehm, Dromey and FURUPS [16], [18], most of them target the software product or process characteristics and does not suit software quality-in-use or require manual user involvement [42], [43]. The McCabe(1976) and Halstead(1977), are used since 1970's while Chidamber& Kemerer metrics[26] triggers its use in 1994. These metrics depend on programming style object-oriented[26] versus procedure programming approaches[24], [25]. Moreover, results from COQUAMO model concluded that there were no software product metrics that were, in general, likely to be good predictors of final product qualities .Thus metrics used in measuring product quality cannot be utilized to measure quality-in-use directly.

## 4. PROPOSED FRAMEWORK

Opinion mining or sentiment analysis is an emerging research direction based on Natural Language Processing that targets to analyze textual user judgments about products or service[11], [12]. Reviews text snippets are good sources for users to decide software purchase and they are a goldmine for product providers. It is obvious that, the average human reader will have difficulty accurately summarizing relevant information and identifying opinions contained in reviews about a product. Moreover, human analysis of textual information is subject to considerable biases resulted from preferences and different understanding of written textual expressions. Therefore, opinion mining provided an alternative to identify important reviews and opinions to answer users' queries[44], [45].

Despite the difficulties of sentiment analysis approach[46], [47][48], it can be used to overcome issues discussed in Section 3. The sentiment analysis can seemingly work on user reviews without active user involvement. Moreover, by sentiment analysis the software trends can be analyzed and future software quality can be predicted.

Next are the details of the framework.

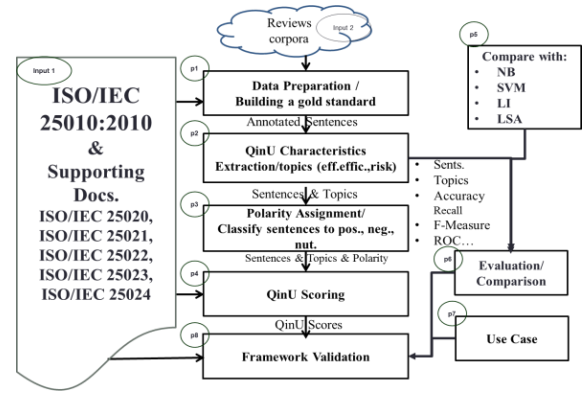

**Fig 3:** Proposed Quality-in-use Prediction Framework

### 4.1 Proposed Quality-in-use Prediction Framework

Fig. 3 shows a general framework to predict software Quality-in-use. We reemphasize that the purpose of this framework is to present a conceptual model to highlight high level details of the proposed framework. In this figure processes are marked by the letter p followed by process number and the input is marked by input1 and input2 respectively. The general idea of the framework is to utilize software reviews and ISO standard documents in order to process review text snippets into QinU characteristics.

The proposed framework has two inputs, the ISO QinU documents (input1) and the software reviews (input2). In this framework, the ISO documents have the quality-in-use description, modeling, specification and evaluation process components. These components will be used to 1) describe QinU for annotators, 2) to get the QinU characteristics, 3) to score QinU using formulas, and 4) to help validate the proposed model. The software reviews text is fed into a data preparation process (p1) in order to get a set of annotated/classified sentences that would be used as a gold standard.

In the data preparation process, first reviews are crawled from software, websites such as Amazon and CNET. These reviews have a star rating from 1 to 5, where 1 stands for bad comment about the software and 5 stands for excellent comment. To balance the input data, for each star rating the top 10 reviews are selected for the next step. This process ensures that the input comments covers the whole star rating range of comments. Next, the reviews from the previous step are split into sentences using a combined automatic method and manual method to cover long sentences or sentences that are missing punctuated. Taking a sentence at a time and within the context of the review the annotator assigns a topic to the sentence. If the sentence is topic related, the annotator will assign the keyword that makes that sentence for a certain topic. For example the annotator might select the word fast from the sentence "this software is fast" as a keyword for the topic efficiency. Additionally the annotator will choose why a sentence is positive or negative by choosing an opinion, expression word and a modifier if available. Finally the classified sentences are saved in the database for the next step which is QinU extraction.

The core of this framework is the QinU extraction. In this step it is proposed to use a sentence semantic similarity measure to map testing sentences into QinU characteristics (p2). In this step the sentences are classified into 3 topics; effectives, efficiency, and risk mitigation using a proposed sentence similarity. In the next step, each sentence is assigned a polarity by using data from the gold standard data set (p1). QinU is scored in process p4 by using linear combination of Qin Characteristics described in the ISO QinU Measurement Standard ISO 25022.

To evaluate the efficiency of the proposed sentence measure in process p2, it can be compared with other famous approaches. Several methods can be used for comparison. This paper chooses to compare with the below methods. The reason for choosing these methods is to have different measures from different spectrums. Li and google tri-grams [49] as sentence measures, Multinomial Naive Bayes text classification (NB) [50]–[52] and SVM[53] for text classification, LSA [54] for semantic space classification. These methods are evaluated in terms of standard classification performance measures: f-measure, accuracy and ROC analysis shown in process p6. In order to validate the framework a Use Case is built in p7 to validate the model.

## 5. PRELIMINARY RESULTS

First the F-measure experiments are shown. Then the top 5 topic words are shown.

### 5.1 F-measur Results

To show the validity of this work, 600 software sentences were labelled to QinU topics. Then 3 algorithms were run; the Multinomial Naive Bayes (NB) Measure algorithm[55], The Multiclass SVM[56], and the Latent semantic analysis[57]. These methods were able to detect a test sentence topic (illustrated in the P2 step of the Fig.3). The experiment was run on 3 fold cross validation. **Fig.** Fig. 4 shows the F-measure of these measures. The figure shows that the sentence length has a direct effect on the final F-measure. Short sentences will tend to have less common words and thus low F-measure.

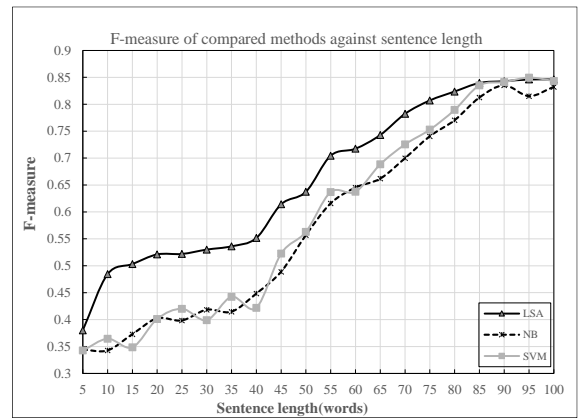

**Fig. 4**.Compared Methods against F-measure

### 5.2 Top Five Keywords

Table 2 shows the top 5 keywords in each topic. From the table, we can see that the words in effectiveness are talking about doing the job, the words in efficiency are talking about expenditure of resources. The risk keywords are talking about the possibility of losing data.

**Table 2:** Top five topic keywords for QinU

| Effectiveness | Efficiency | Freedom from Risk |
|---|---|---|
| work | speed | issue |
| features | stable | trouble |
| interface | slow | error |
| simple | load | Freeze |
| easy | memory | fix |

## 6. RELATED WORKS

Feature or topic extraction has been discussed in literature in many works such as [48], [58]–[61]. Most of these works use the language semantics to extract features such as nouns and noun phrases along with their frequencies subject to predefined thresholds. Qiu et al. [45], [61] suggested to extract both features and opinion by propagating information between them using grammatical syntactic relations.

Leopairote, Surarerks, and Prompoon[10] proposed a model that can extract and summarize software reviews in order to predict software "quality-in-use". The model depends on a manually built ontology of ISO 9126 "quality-in-use" keywords and Word Net 3.0 synonyms expansion. We consider the work of [10] the most nearby to this paper.

The difference from proposed work is that the proposed framework employs word similarity and relatedness rather than rule based classification and ontologies.

## 7. CONCLUSION

Quality-in-use represents software quality in the viewpoint of a user. This paper presents the major issues in measuring software quality-in-use. Quality-in-use can be measured using standard SQuaRE series while many characteristics of software quality-in-use are scattered in many customized software quality models. Measuring quality-in-use is challenging, due to the complexity of current standard models and the incompleteness of other related customized models. The viewpoint of the software users is hard to be implemented within the software lifecycle ahead of time especially for hedonic tasks. This paper proposes to process software reviews in order to get software quality-in-use. The framework employs sentence semantic relatedness to get a score for QinU characteristics.

## 8. ACKNOWLEDGMENT

This study was supported in part by University Malaysia Sarawak' Zamalah Graduate Scholarship and grant ERGS/ICT07 (01) /1018/2013 (15) Minister of Education Malaysia.